# Selective coupling of coherent phonons to intertwined charge-orbital and magnetic orders in doped manganites

X. Liu[1*], M. Sander[1], S.-W. Huang[1], S. Zerdane[1], A. Caviezel[1], M. Rössle[2,3], J. Lu[4], D. Babich[1], S. Shin[5], E. Pomjakushina[5], P. Marsik[2], L. Wang[4], S. W. Cheong[6,7], C. Jia[8], P. Beaud[1], H.T. Lemke[1], U. Staub[1], R. Mankowsky[1*]

[1]Center for Photon Science, Paul Scherrer Institute, 5232 Villigen PSI, Switzerland
[2]Department of Physics, University of Fribourg, 1700 Fribourg, Switzerland
[3]Helmholtz-Zentrum Berlin für Materialien und Energie GmbH, Wilhelm-Conrad-Röntgen Campus, BESSY II, 12489 Berlin, Germany
[4]Hefei National Research Center for Physical Sciences at Microscale, University of Science and Technology of China, 230026 Hefei, China
[5]Center for Neutron and Muon Sciences, Paul Scherrer Institute, 5232 Villigen PSI, Switzerland
[6]Keck Center for Quantum Magnetism, Rutgers University, Piscataway, 08854 New Jersey, USA
[7]Department of Physics and Astronomy, Rutgers University, Piscataway, 08854 New Jersey, USA
[8]Key Laboratory for Magnetism and Magnetic Materials of the Ministry of Education and Lanzhou Center for Theoretical Physics, Lanzhou University, 73000, Lanzhou, China

[*]Corresponding authors.
Email: xin.liu@tuwien.ac.at; roman.mankowsky@psi.ch

**Abstract**
Strongly correlated materials feature technologically relevant functionalities such as high-temperature superconductivity and colossal magnetoresistance, which emerges from the competition and coexistence of electronic and magnetic phases. Uncovering the microscopic interactions underlying these phenomena remains challenging because spin, orbital, charge, and lattice degrees of freedom are inherently intertwined. Here, by combining time-resolved X-ray diffraction and polarization-resolved ultrafast optical reflectivity in $La_{1/4}Pr_{3/8}Ca_{3/8}MnO_3$, we reveal that coherent phonon modes can selectively track different ordered phases: the in-plane phonon response is predominantly sensitive to charge/orbital order, and the *c*-axis phonon response is sensitive to magnetic order. Moreover, a ferromagnetic-related hysteresis emerges even when solely probing the charge/orbital-ordered phase, indicating a strong microscopic coupling between the spatially separated charge/orbital-ordered and ferromagnetic-ordered phases. These results demonstrate that coherent phonons provide a direct time-domain route to disentangle intertwined electronic and magnetic dynamics in coupled phases.

## INTRODUCTION

The intricate interplay between spin, lattice, charge and orbital degrees of freedom plays a vital role in various emergent phenomena and functionalities in strongly correlated electron systems, including colossal magnetoresistance (CMR), unconventional superconductivity, multiferroicity and topological orders/phases *(1-6)*. Capturing the dynamics of these intertwined orders is critical not only for elucidating the fundamental physics behind the emergent phenomena in correlated systems, but also for advancing their use in next-generation ultrafast electronic and spintronic devices *(1,2,7,8)*. Coherent phonons, as collective atomic vibrations, offer a unique opportunity to probe and manipulate these functionalities as reported in prior studies *(9-13)*. In addition, these specific phonon modes could also strongly couple with electronic or magnetic orders in correlated materials, representative examples include charge/orbital and spin orders *(10,13-15)*. However, direct experimental evidence for how such intertwined orders selectively couple to anisotropic coherent phonon responses at ultrafast timescales remains scarce, especially in systems exhibiting phase coexistence or multiple competing ground states.

$La_{1/4}Pr_{3/8}Ca_{3/8}MnO_3$ (LPCMO) provides an ideal platform for investigating such coupling among the competing order parameters. As a mixed-valence manganite ($Mn^{3+}$ and $Mn^{4+}$) and one of the most extensively studied CMR material *(16-19)*, LPCMO exhibits rich phases and orders shown in Fig. 1, which are characterized by coexisting and competing charge, orbital, spin, and lattice orders *(20-22)*. At room temperature, LPCMO is a paramagnetic insulator with an orthorhombic lattice structure (*pbnm*). Upon cooling, the system exhibits a charge order (CO) and orbital order (OO) phase transition at ~220 K ($T_{co/oo}$). During the transition, the $Mn^{3+}$ and $Mn^{4+}$ ions arrange themselves in an ordered checkerboard-like pattern within the lattice, while the $d_{3x^2-r^2}$ and $d_{3y^2-r^2}$ orbitals align in a zig-zag pattern, forming the orbital order *(23)*. In addition, a structural change from the orthorhombic to a monoclinic ($P2_1/m$) phase occurs *(17,23)*. Upon further cooling to 180 K ($T_N$), antiferromagnetic order emerges, followed by a ferromagnetic phase transition and insulator-to-metal transition at ~80 K ($T_C$) *(17,24)*. All these phases coexist in a temperature-dependent ratio, giving rise a ground state with coexisting ferromagnetic metallic and CO/OO antiferromagnetic insulating phases.

In this work, we disentangle the ultrafast dynamics of the charge/orbital and spin orders in LPCMO single crystals by combining time-resolved X-ray diffraction (XRD) and polarization-resolved optical pump-probe reflectivity. We found that the coherent phonon responses exhibit strong directional selectivity: they are predominantly sensitive to charge/orbital order within the *ab*-plane, and to spin order along the *c*-axis. The later, spin-coupled mode is attributed to the enhanced out-of-plane $d_{3z^2-r^2}$-like character of the photoexcited states associated with $Mn^{3+}$-$Mn^{4+}$ charge-transfer excitations along the zig-zag chains in the antiferromagnetic background. This phonon frequency undergoes further hardening when passing into the ferromagnetic phase. In addition, a ferromagnetic-related hysteresis behavior was observed in the charge/orbital phases, identifying the microscopic coupling between the spatially separated charge/orbital and spin ordered phases in LPCMO.

## RESULTS

### Temperature evolutions of phases in $La_{1/4}Pr_{3/8}Ca_{3/8}MnO_3$

Phenomenologically, following related treatments of order parameter-phonon coupling *(14,15)*, the coupling of phonons to CO/OO and spin orders can be described by the effective Landau-type phonon free energy:

$$F_{\mathrm{ph}} = \frac{1}{2}\omega_1^2 Q_1^2 + \frac{1}{2}\omega_2^2 Q_2^2 + \frac{1}{2}\lambda_{CO/OO}\eta^2 Q_1^2 + \frac{1}{2}\lambda'_{CO/OO}\eta^2 Q_2^2 + \frac{1}{2}\lambda_m m^2 Q_1^2 + \frac{1}{2}\lambda'_m m^2 Q_2^2. \quad (1)$$

Here $Q_1$ and $Q_2$ are phonon modes, $\omega_1$ and $\omega_2$ are their frequencies, $\eta$ and $m$ are CO/OO and spin order parameters, and $\lambda_{CO/OO}$, $\lambda'_{CO/OO}$, $\lambda_m$ and $\lambda'_m$ denote the coupling strengths of the phonons to the CO/OO and spin orders, respectively. In the LPCMO single crystal with an (100) surface normal, the presence of these multiple intertwined orders was firstly confirmed by measuring the temperature dependent magnetization (Materials and Methods). As shown in Fig. 1, three phase transitions were observed, in agreement with previous studies mentioned above *(17,24)*. In addition, a hysteresis behavior below 100 K was seen during the cooling and warming cycles, consistent with previously reported behavior *(19,21)*. This can be interpreted as the result of a phase transition between ferromagnetic and charge/orbital ordered antiferromagnetic phases, leading to a path-dependent evolution of the competing orders during cooling and warming *(19)*. Furthermore, temperature dependent resonant XRD studies were performed to investigate the phase coexistence in the sample. Resonant XRD enables direct detection of charge, orbital, or spin order by tuning the X-ray energy to an atomic absorption edge, selectively enhancing scattering from specific electronic configurations. In LPCMO (*Pbnm* with $k$ odd): (0 $k$/2 0) is sensitive directly to the doubling of the lattice constant in the orbital ordered phase; (0 $k$ 0) is more sensitive to the charge order *(17,19)*. As shown in Fig. S1 and S2, the (0 3 0) charge order peak shows up at 220 K with gradually increasing intensity during cooling. At 80 K, the intensity of the charge order peak starts to decrease, due to the formation and growth of the ferromagnetic phase, which is not charge ordered. This resonant XRD result is consistent with the magnetization measurement, showing the competition between electronic and magnetic orders in LPCMO.

**Optical anisotropy**

Coherent phonons can be created by ultrafast laser excitations *(25,26)*. To identify a suitable wavelength of the laser excitation, the optical conductivity in LPCMO single crystal was characterized by using spectroscopic ellipsometry (Materials and Methods). As shown in Fig. 2a and 2b, the optical conductivity exhibits pronounced anisotropy for light polarized along the $b$ and $c$ axes (E // $b$ in Fig. 2a; E // $c$ in Fig. 2b). In the spectra, two main optical bands were observed: a high-energy band centered around 4 eV, assigned to the charge transfer between O 2$p$ to Mn 3$d$ orbitals *(27)*, and a low-energy band below 2 eV, where the most prominent anisotropic features appear. To quantify this anisotropy, the spectral weight (SW) below 2 eV was calculated. As plotted in Fig. 2c, the normalized SW along E // $b$ sharply increases below ~220 K, linked with the onset of charge and orbital orders. In contrast, the SW along E // $c$ increases near 180 K, coinciding with the antiferromagnetic transition. A closer inspection of the E // $c$ spectra shows two contributions to the sub-2 eV band, labeled A (below 1 eV) and B (above 1 eV) in the inset of Fig. 2b. Similar features have also been observed in other manganites *(27,28)*: peak B is typically associated with a transition between the Jahn-Teller (JT) spilt $e_g$ bands of $Mn^{3+}$ ions; peak A arises from an interatomic $e_g$-electron transition from the JT distorted $Mn^{3+}$ to the $Mn^{4+}$ ions (polaron hopping). The temperature dependence of the normalized SW for these two components is shown in Fig. 2d, peak B is independent as a function of temperature, whereas peak A shows a clear increase below $T_N$. All these optical conductivity results suggest that the low-energy charge-transfer spectral weight in LPCMO, associated with broad $Mn^{3+}$-$Mn^{4+}$ intersite $e_g$-electron excitations, is strongly influenced by both charge/orbital and spin orders, leading to a pronounced anisotropy along different crystallographic directions.

**Visualization of selective coherent phonon coupling to charge-orbital and spin orders**

Based on the low-energy electronic features revealed by the optical conductivity in Fig. 2, an excitation source of 800 nm (~1.55 eV, as the dashed line shown in Fig. 2a and 2b) femtosecond laser pulse was selected to conduct time-resolved optical reflectivity and XRD measurements. This photon energy falls within the broad absorption region and allows to efficiently excite various phonon modes in manganite compounds coherently by a displacive excitation mechanism *(29,30)*. Figure 3a shows the schematic of the optical pump probe reflectivity measurement, where the polarization of the probe beam (800 nm) was rotated either along *b* (E // *b*) or *c* (E // *c*) axis. At 6 K, for both cases (E // *b* and E // *c*) with a fluence of 1 mJ/cm$^2$, oscillations were observed in the time trace of reflectivity (Fig. 3b), which originate from a coherent lattice response *(29,30)*. Notably, at this low fluence, the observed oscillations serve as a sensitive probe without melting any phases *(30)*. Temperature dependent measurements reveal a pronounced anisotropy in the response along the two directions (see Fig. S3 and S4 for individual traces). For polarization along the zig-zag chains E // *b*, the oscillation disappears at ~220 K (Fig. 3c and S3c) near the CO/OO phase transition, while the oscillation disappears at ~180 K (Fig. 3d and S3d) near the antiferromagnetic phase transition for perpendicular polarization E // *c*. In addition, at 80 K near the transition temperature of the ferromagnetic phase, there is a subtle change of the oscillation only for E // *c* (Fig. S3d). To extract the temperature dependent frequency of the oscillation, we fit the optical pump probe reflectivity (black curve shown in Fig. 3b) by combining an error function, exponential decay and oscillation functions (see Materials and Methods). As seen in Fig. 3e, a frequency of 2.45 THz was obtained for E // *b* (blue), which remains almost from 6 K to 220 K. In contrast, the oscillations for E // *c* (red) exhibit a higher frequency of 2.6 THz at 6 K, which softens to ~2.5 THz at the ferromagnetic phase transition and disappears above ~180 K at the AFM phase transition. Furthermore, a hysteresis behavior consistent with the magnetization measurement was observed for E // *c* during cooling and warming cycles (see Fig. 3f). These results demonstrate that the coherent phonon responses are selectively coupled to the CO/OO and spin orders in LPCMO for polarization along *b* and *c*, respectively.

Although the optical pump probe measurements reveal clear direction-dependent coherent phonon oscillations and provide evidence for selective coupling to CO/OO and spin orders, these measurements average over all phases simultaneously in LPCMO and the reflectivity changes alone cannot unambiguously be attributed to a specific phase of LPCMO. To disentangle the structural response of the phases, we further performed time-resolved XRD measurements at Bernina beamline in SwissFEL *(31-34)* (Materials and Methods). The doubling of the crystal *b* lattice constant in the CO/OO ordered phase gives rise to superlattice peaks with half integer of *k*, which are exclusively sensitive to the CO/OO phase *(19,24)*. By comparing measurements of the (2 -3.5 0) peak, which is insensitive to structural changes along the *c* axis as its *l* component is 0, and the (2 0.5 -4) peak with large *l* and therefore high sensitivity to atomic motions along the *c* axis, we can further resolve the dynamics along the different crystal directions. The geometry of the measurements is schematically shown in Fig. 4a.

Figure 4b shows time dependence of the relative changes of the (2 -3.5 0) and (2 0.5 -4) reflection intensities upon excitation with 0.7 mJ/cm$^2$ 800 nm pulses at 20 K. Fitting of the observed oscillations yields frequencies of 2.45 THz and 2.6 THz, respectively. Temperature dependent measurements were also performed (see Fig. S5). As shown in Fig. 4c, the frequency of the (2 0.5 -4) peak exhibits similar softening for increasing temperatures as observed in optical pump probe measurements with probe polarization along the *c* axis. Surprisingly, both techniques reveal a pronounced hysteresis behavior (Fig. 3f and 4d). As the reflection intensity is solely sensitive to the CO/OO phase, the appearance of the hysteresis in the coherent mode frequency collected in XRD (Fig. 4d) indicates a strong coupling between the spatially separated CO/OO and ferromagnetic phases in LPCMO. This may also suggest that the two phases are not independent, but coexist as nanoscale or mesoscopic domains that remain coupled through interfacial strain and exchange interactions. The hysteresis of the ferromagnetic phase can therefore modify the local lattice potential of neighboring CO/OO regions, imprinting a ferromagnetic-related hysteresis onto the CO/OO-sensitive coherent phonon response. The time-resolved XRD results reveal clear oscillations in the diffraction intensity of both (2 -3.5 0) and (2 0.5 -4) superlattice peaks, with frequencies matching those observed in the optical reflectivity signals. This confirms that the coherent phonon modes originate from real-space lattice displacements and not solely from optical modulation effects *(35)*. The (2 -3.5 0) peak without sensitivity to ionic motions along the *c* axis exhibits a dominant 2.45 THz oscillation, which is consistent with the CO/OO order; While the (2 0.5 -4) peak, sensitive also to structural distortions along the *c* axis, shows a 2.5 THz (frequency hardening in the ferromagnetic state to 2.6 THz) that evolves with AFM spin order. These findings reinforce the interpretation of symmetry-selective phonon coupling to CO/OO and spin orders.

## DISCUSSION

Motivated by the earlier Landau-type descriptions of order parameter-phonon coupling and by our experimental observations, we introduce the following minimal effective phonon free energy, in which the in-plane phonon mode $Q_b$ couples predominantly to the CO/OO order parameter $\eta$, while the out-of-plane-sensitive mode $Q_c$ couples predominantly to the antiferromagnetic and ferromagnetic spin order parameters $m_{AFM}$ and $m_{FM}$:

$$F_{\mathrm{ph}} = \frac{1}{2}\left(\omega_b^2 + \lambda_{CO/OO-ph}\eta^2\right)Q_b^2 + \frac{1}{2}\left(\omega_c^2 + \lambda_{AFM-ph}m_{AFM}^2 + \lambda_{FM-ph}m_{FM}^2\right)Q_c^2. \quad (2)$$

Here $\omega_b$ and $\omega_c$ are the phonon frequencies, and $\lambda_{CO/OO-ph}$ and $\lambda_{AFM-ph}/\lambda_{FM-ph}$ denote the coupling strengths of the phonons to the CO/OO and antiferromagnetic/ferromagnetic spin order parameters, respectively. The cross-coupling terms shown in Eq. (1) are neglected here since no such signatures appear in the experimental data. In this phenomenological description, the effective phonon frequencies are renormalized as

$$\Omega_b^2 = \omega_b^2 + \lambda_{CO/OO-ph}\eta^2, \;\; \Omega_c^2 = \omega_c^2 + \lambda_{AFM-ph}m_{AFM}^2 + \lambda_{FM-ph}m_{FM}^2. \quad (3)$$

This form captures the experimental observation that the $b$-axis-sensitive mode is mainly governed by the CO/OO, whereas the $c$-axis-sensitive mode is predominantly coupled to spin orders.

In the CO/OO phase of LPCMO, electrons on $Mn^{3+}$ sites alternatively occupy the $d_{3x^2-r^2}$ and $d_{3y^2-r^2}$ orbitals *(23)*. This orbital configuration is accompanied by static JT-related lattice distortions, as commonly found in similar manganites *(30,36,37)*. Ultrafast photoexcitation primarily perturbs the in-plane $Mn^{3+}_{3x^2-r^2/3y^2-r^2}Mn^{4+} \rightarrow Mn^{4+}Mn^{3+}_{3x^2-r^2/3y^2-r^2}$ charge-transfer network along the zig-zag chains *(38-40)*, which suddenly reshapes the lattice potential associated with the CO/OO background and coherently drives the 2.45 THz mode (Fig. 5). This mode is therefore assigned to a low-frequency structural response, predominantly coupled to the in-plane charge/orbital texture. Upon entering the antiferromagnetic phase, the pronounced enhancement of the E // *c* optical conductivity (Fig. 2c) suggests that the corresponding low-energy optical transitions acquire a stronger out-of-plane character, consistent with polarization-dependent optical studies of orbital-ordered manganites showing that the optical spectral weight is strongly governed by orbital-dependent transition matrix elements *(41,42)*. This enhancement can reflect an enhanced $d_{3z^2-r^2}$-like admixture ($Mn^{3+}_{3x^2-r^2/3y^2-r^2}Mn^{4+} \rightarrow Mn^{4+}Mn^{3+}_{3z^2-r^2}$) in the relevant photoexcited states *(39)* (Fig. 5). This orbital reweighting strengthens the coupling to a $c$-axis-sensitive lattice coordinate and activates the 2.5 THz coherent phonon. Upon further cooling below the ferromagnetic transition, the enhanced ferromagnetic correlations can further renormalize the effective force constant of the spin-coupled mode through spin-phonon coupling and exchange-striction effects, leading to its hardening to ~2.6 THz. This is consistent with anomalous phonon-frequency renormalization observed in manganites across magnetic ordering transitions *(43-45)*. In contrast, the 2.45 THz mode remains nearly unchanged because it is primarily governed by the CO/OO-related in-plane lattice potential. The estimated ferromagnetic-spin-phonon coupling constant is ~2.68 $cm^{-1}$ (Fig. S6), which is comparable to values reported for other manganites *(43,44)*. This interpretation is further supported by the fluence-dependent ultrafast optical reflectivity and XRD measurements (Figs. S7 and S8): increasing pump fluence induces a progressive redshift of the 2.6 THz mode (Fig. S9), consistent with a melting of magnetic order at high excitation density, whereas the 2.45 THz mode remains largely fluence-independent (Fig. S9), in line with its dominant coupling to the more robust CO/OO-related lattice distortion *(46,47)*.

In summary, we have uncovered direction-dependent coupling of coherent phonons to charge/orbital and spin orders in LPCMO at ultrafast timescales. The distinct frequency evolution across multiple phase transitions reveals the central role of lattice dynamics in mediating the interplay among charge, orbital, and spin degrees of freedom. Our results provide direct time-domain evidence that spin-phonon and electron-phonon coupling is governed by crystallographic anisotropy and phase competition. This work further highlights that anisotropic coherent phonons as order-sensitive probes offer a potential route to disentangle and manipulate intertwined orders in correlated materials.

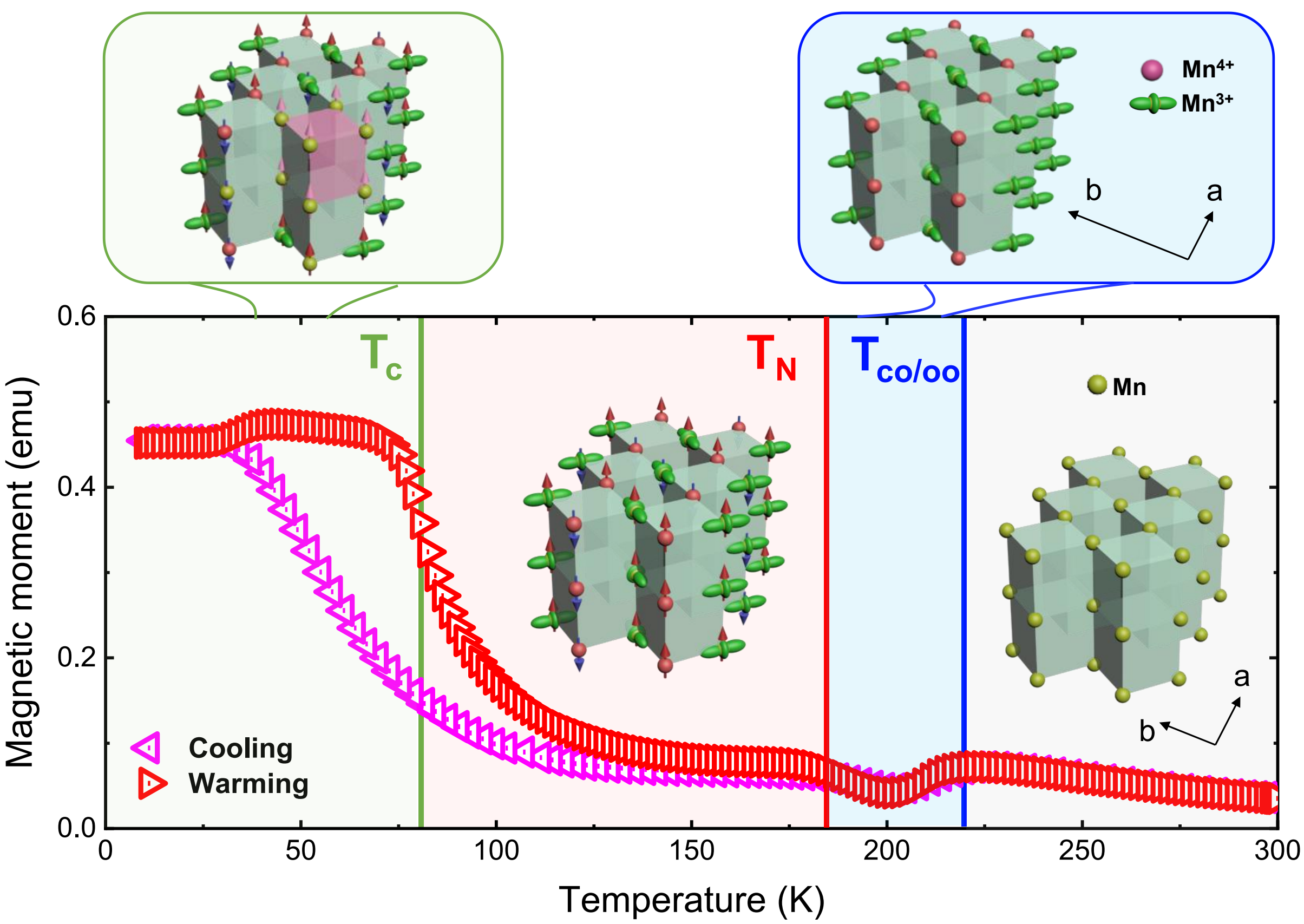

**Fig. 1. Temperature evolutions of phases in $La_{1/4}Pr_{3/8}Ca_{3/8}MnO_3$ (LPCMO).** Magnetization as a function of temperature shows the various phase transitions. T>220 K: paramagnetic insulating phase with an orthorhombic lattice structure; T=220 K ($T_{co/oo}$): charge/orbital order phase transition accompanying with a structural transition to monoclinic, where the lattice constant along *b* axis will be doubled; T=180 K ($T_N$): antiferromagnetic phase transition; T=80 K ($T_C$): ferromagnetic phase transition, where the ferromagnetic metallic phase will coexist with the charge/orbital antiferromagnetic phase. The insets in each temperature range show the schematic lattice and spin structures. The pink and red triangles are measured during cooling and warming cycles under a magnetic field of 0.2 T, respectively. A hysteresis behavior was observed around the ferromagnetic phase transition.

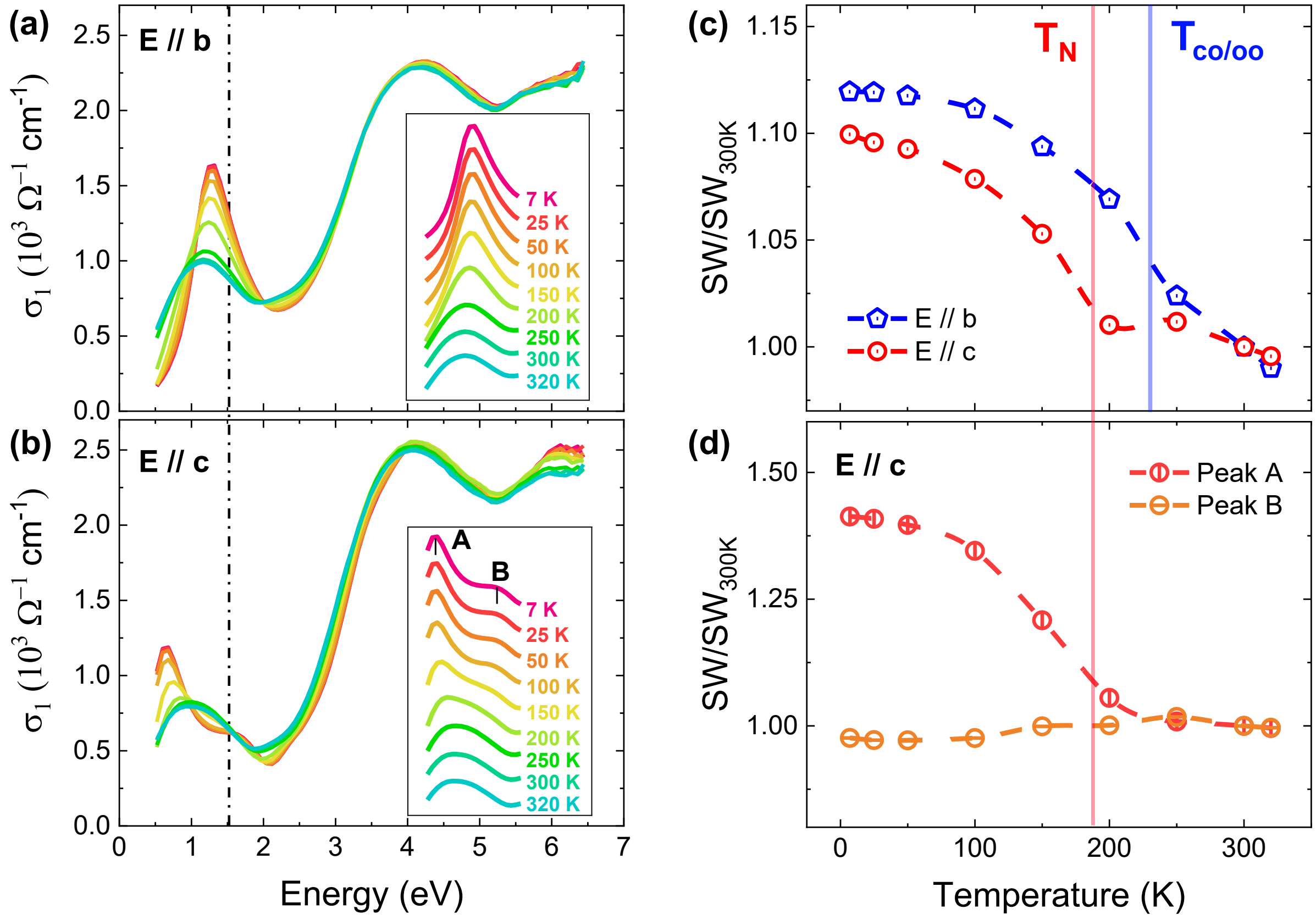

**Fig. 2. Optical anisotropy in LPCMO.** (a)(b) Temperature dependent optical conductivity in LPCMO. Polarization along *b* axis (E // *b*) in (a) and polarization along *c* axis (E // *c*) in (b). Insets show the zoom-in range below 2 eV. The black dashed line indicates the energy of excitation by 800 nm laser pulse. (c) Spectral weight (SW) below 2 eV as a function of temperature. The blue open pentagons and red open circles are polarization along *b* axis and *c* axis, respectively. The SW was normalized by SW at 300 K ($SW_{300K}$). (d) Normalized SW for polarization along *c* axis, where the peak A (red) and peak B (orange) are as labeled in (b). An anisotropic behavior was observed for the polarization along different axis, corresponding to different transition temperatures.

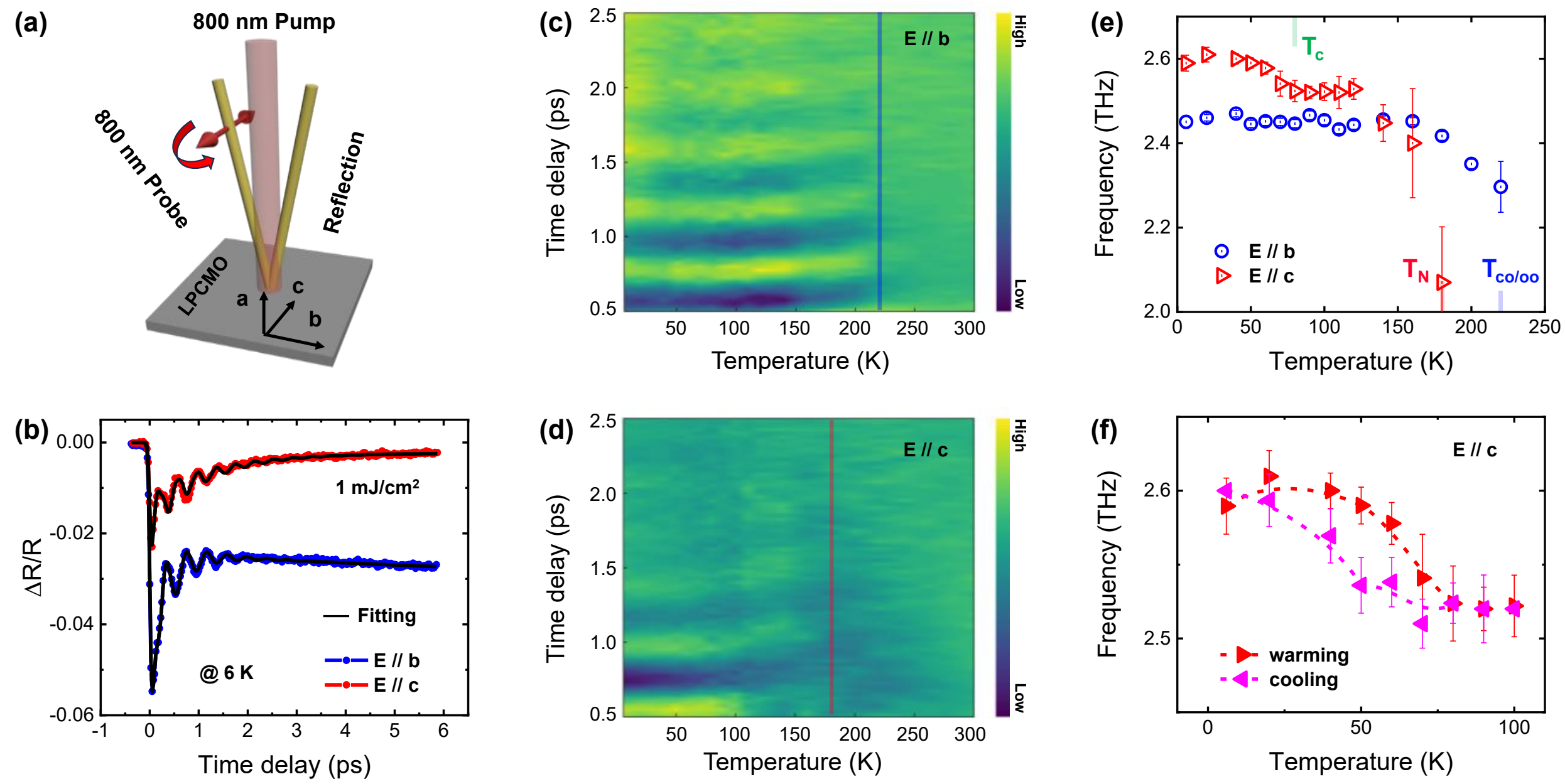

**Fig. 3. Selective coherent phonon coupling to charge-orbital and spin orders in LPCMO via ultrafast reflectivity.** (a) The schematic of the optical pump probe reflectivity measurement using 35 fs, 800 nm pulses. (b) Time trace of the reflectivity at 6 K with 1 mJ/cm$^2$ fluence, where the polarization of the probe beam is along *b* (blue, E // *b*) or along *c* (red, E // *c*) axis, respectively. (c)(d) Temperature-time-delay contour maps of the extracted oscillations for E // *b* and E // *c*, respectively. (e) Frequency as a function of temperature for E // b (blue open circles) and E // *c* (red open triangles) with the fluence of 1 mJ/cm$^2$. (e) Hysteresis behavior for E // *c* during cooling (pink full triangles) and warming (red full triangles) process.

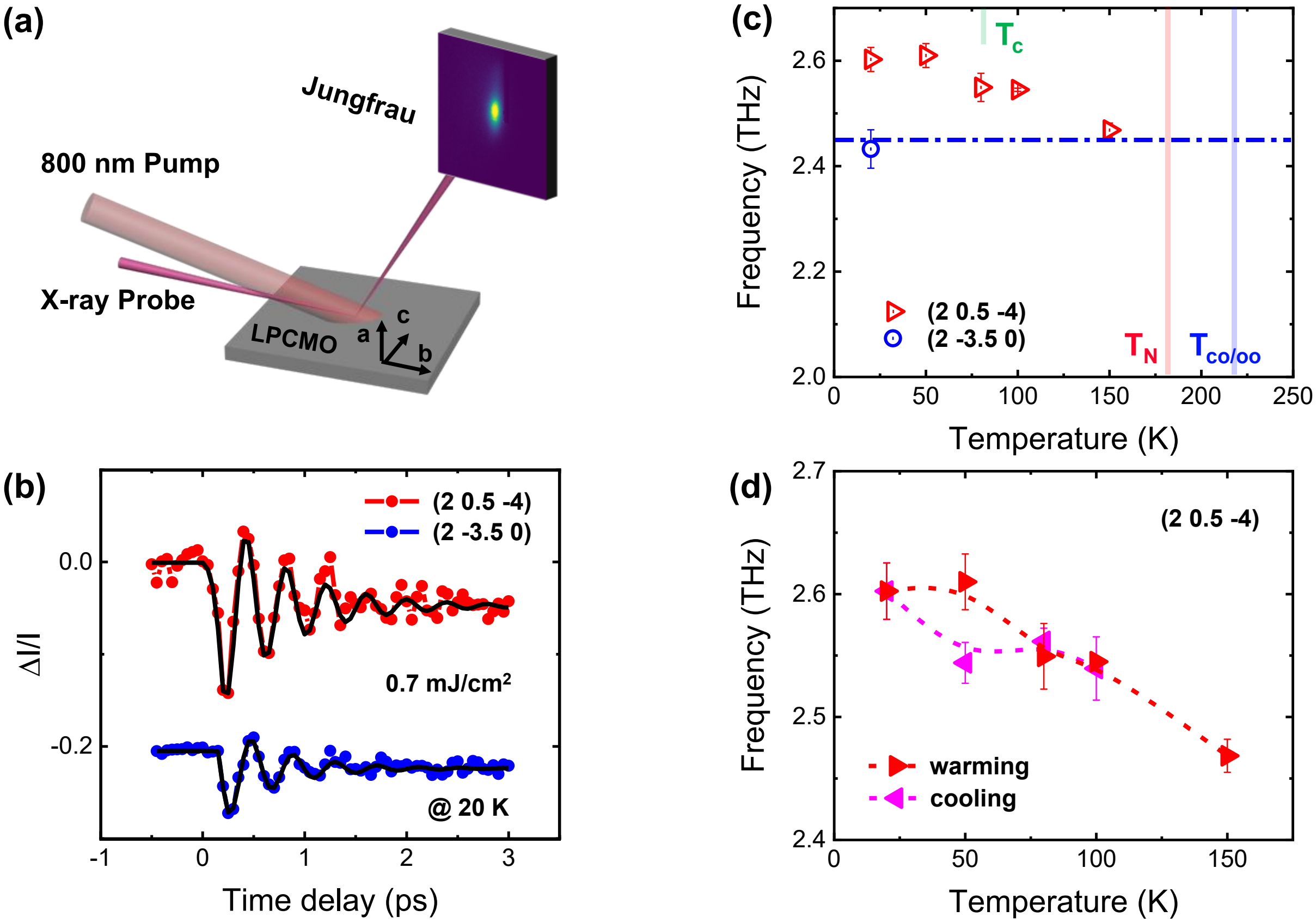

**Fig. 4. Selective coherent phonon coupling to charge-orbital and spin orders in LPCMO via time-resolved X-ray diffraction.** (a) The schematic of the time-resolved X-ray diffraction measurement. 800 nm laser was used as pump beam. (b) Time trace of the intensity of (2 0.5 -4) (red) and (2 -3.5 0) (blue) superlattice peaks at 20 K with the fluence of 0.7 mJ/cm$^2$. (c) Frequency as a function of temperature extracted from the (2 0.5 -4) superlattice peak (red open triangles). The blue open circle is the frequency extracted from the (2 -3.5 0) superlattice peak at 20 K. (d) Hysteresis behavior in frequency from the (2 0.5 -4) peak during cooling (pink full triangles) and warming (red full triangles) process.

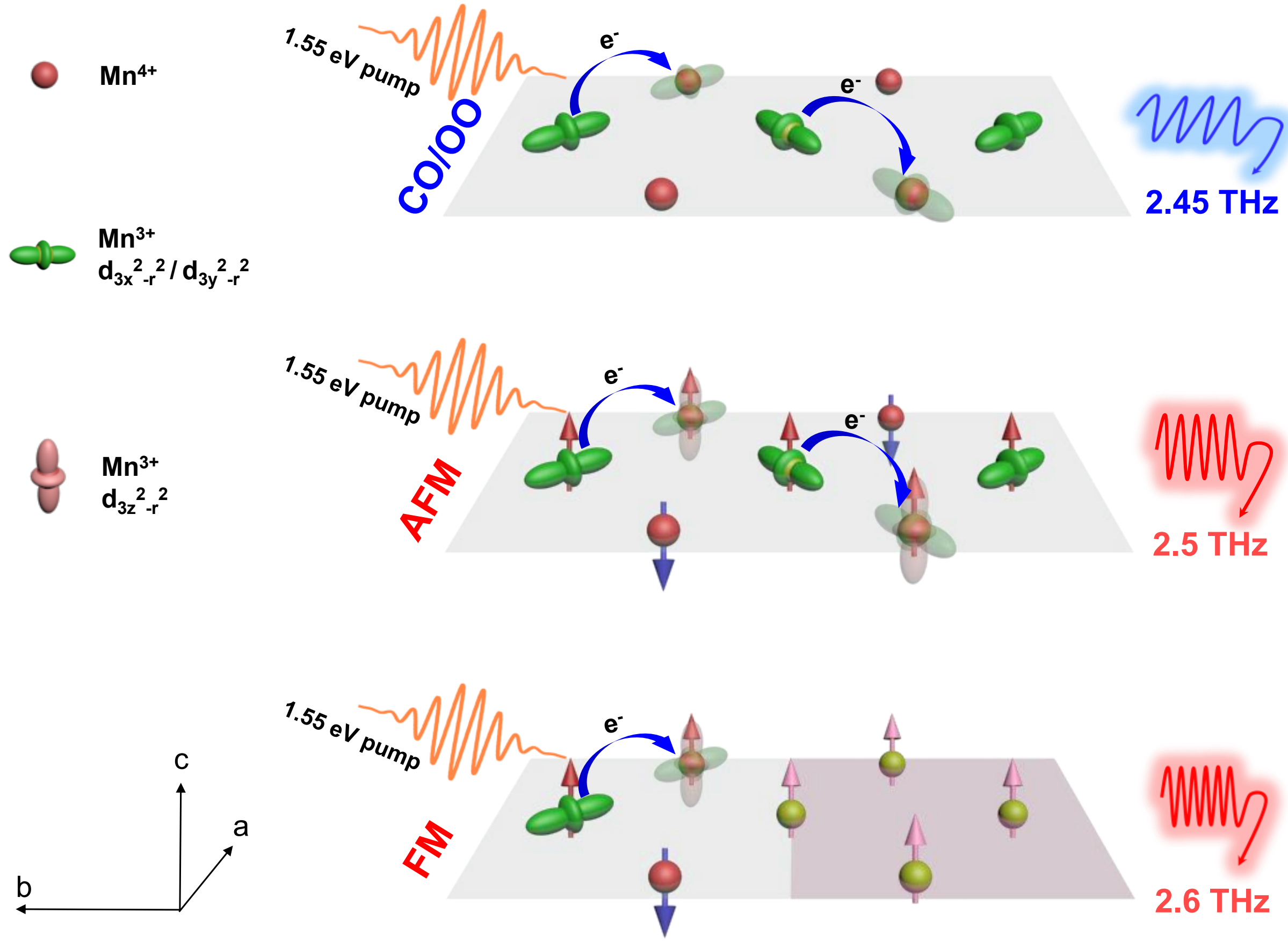

**Fig. 5. Schematic illustration of electronic excitation channels and coherent phonon responses in LPCMO.** T<$T_{\text{co}}$, the $Mn^{3+}$/$Mn^{4+}$ arrangement is accompanied by alternating occupation of the $d_{3x^2-r^2}$ and $d_{3y^2-r^2}$ orbitals. Ultrafast excitation mainly perturbs the in-plane $Mn^{3+}$ - $Mn^{4+}$ electron hopping channel (blue arrows), coherently driving the 2.45 THz phonon mode selectively coupled to the CO/OO order. T<$T_{\text{N}}$, the antiferromagnetic background enhances the $d_{3z^2-r^2}$-like admixture of the relevant photoexcited states, thereby activating a $c$-axis-sensitive 2.5 THz coherent phonon mode. T<$T_{\text{C}}$, additional magnetic renormalization in the ferromagnetic state hardens this mode to ~2.6 THz.

## MATERIALS AND METHODS

### Magnetization

Magnetization as a function of temperature was measured by a SQUID magnetometer (Quantum Design). A magnetic field of 0.2 T was applied during both cooling and warming process to determine the magnetization.

### Ellipsometry

We measured the spectroscopic ellipsometry by using a Woollam VASE ellipsometer to obtain the optical conductivity of the sample. The energy range covers from 0.5 to 6.5 eV. The polarization was fixed either along the $b$ or $c$ axis of the sample during the measurements. The LPCMO single crystal was mounted in a helium flow cryostat (CryoVac) under ultrahigh vacuum, allowing measurements from room temperature down to 7 K.

### Resonant X-ray diffraction

The temperature dependence of charge order peak (0 3 0) was measured at MS beamline (X04SA) of Swiss Light Source at the Mn $K$ edge of ~6.55 keV. The sample was rotated during the measurement to access the peak.

### Optical pump probe reflectivity

Optical pump probe reflectivity measurements were carried out with a 1 kHz Ti:sapphire laser system at a wavelength of 800 nm. The beam was split by beam splitter into pump and probe beams, then two beams were focused on the sample with the spot size of 350μm × 380μm (pump) and 200μm × 180μm (probe), respectively. A cross-polarization setup between the pump and probe beam was used to eliminate interferences. The polarization of the probe beam was rotated either along $b$ or $c$ axis to measure the reflectivity (R). The experimental set-up is depicted schematically in Fig. 3a. Near normal incident geometry was used. In the experiment, a closed cycle refrigerator (Janis RDK-205D) was used to obtain temperatures down to 6 K.

To fit the results of $\Delta R/R$, a combined function was used, which includes one error function, bi-exponential decay function and two oscillation functions:

$$\frac{\Delta R}{R(t)} = \mathrm{erf}_{step(t,t_0,\sigma)} \cdot \left[ \sum_{i=0}^{1} A_\mathrm{i}\, e^{-\frac{t-t_0}{\tau_\mathrm{i}}} + \sum_{j=2}^{3} A_\mathrm{j}\, e^{-\frac{t-t_0}{\tau_\mathrm{j}}} \cos\left(2\pi \cdot (\omega_\mathrm{j}(t-t_0) + \Phi_\mathrm{j})\right) + Y_0 \right],$$

where $\mathrm{erf}_step(t, t_0, \sigma) = \frac{1}{2}(1 + \mathrm{erf}(\frac{t-t_0}{\sigma}))$ represents a smooth step function centered at $t_0$ with a characteristic rise time governed by $\sigma$; $t$ represents the time delay; $A_0$ and $A_1$ are amplitudes of exponential decay components with corresponding time constants $\tau_0$ and $\tau_1$; $A_2$ and $A_3$ are amplitudes of damped oscillatory components with decay constants $\tau_2$ and $\tau_3$, frequencies $\omega_2$ and $\omega_3$, phases $\Phi_2$ and $\Phi_3$; $Y_0$ is a constant offset. The fitted frequency $\omega_2$ was shown in Fig. 3e. While the other frequency $\omega_3$ is around ~6 THz and temperature independent, as reported in the literature *(29)*.

### Time-resolved X-ray diffraction

Time-resolved X-ray diffraction measurements were performed at the SwissFEL Bernina instrument. The experimental set-up is depicted schematically in Fig. 4a. The photon energy was selected near the Mn $K$ edge to measure the superlattice peaks. The X-ray beam was focused to the sample with a grazing incidence angle of ~0.5°. The 800 nm pump laser beam was focused to a spot of 390μm × 380μm with an incidence angle of ~10° with $p$-polarization. The sample was cooled with a helium flow cryostat to a base temperature of 20 K. The femtosecond optical pump laser was synchronized with an X-ray probe to capture transient lattice dynamics.

To fit the results of $\Delta I/I$, where $I$ is the intensity of the peak. A combined function was used, which includes one error function, one exponential decay function and one oscillation function:

$$\frac{\Delta I}{I(t)} = \mathrm{erf}_{step(t,t_0,\sigma)} \cdot \left[ B_1 e^{-\frac{t-t^0}{\tau_1^*}} + B_2 e^{-\frac{t-t^0}{\tau_2^*}} \cos\left(2\pi \cdot (\omega(t-t_0) + \phi)\right) + y_0 \right],$$

where erf_$step(t, t_0, \sigma) = \frac{1}{2}(1 + \mathrm{erf}(\frac{t-t_0}{\sigma}))$ represents a smooth step function centered at $t_0$ with a characteristic rise time governed by $\sigma$; $t$ represents the time delay; $B_1$ is the amplitude of exponential decay with corresponding time constant $\tau_1^*$; $B_2$ is the amplitude of damped oscillatory with decay constants $\tau_2^*$, frequencies $\omega$, phase $\phi$; $y_0$ is a constant offset. The fitted frequency $\omega$ was shown in Fig. 4c.

**Charge transfer analysis**

Suppose the applied optical field is a linearly polarized, monochromatic plane-wave described by

$$\boldsymbol{A}(\omega, r) = A_0 \boldsymbol{\epsilon} e^{i\boldsymbol{k}\cdot\boldsymbol{r} - i\omega t}$$

where $\boldsymbol{\epsilon}$ is the polarization vector, $A_0$ is the amplitude, and $\omega$ is the frequency of the optical field. According to the Fermi "golden rule", the rate from an electron initially in a state $|\psi_i\rangle$ to absorb energy $\hbar\omega$ and transition to a higher energy state $|\psi_f\rangle$ is given by *(48,49)*

$$W_{i\to f} = \frac{2\pi}{\hbar} \sum_f \left|\langle \psi_f | V(\boldsymbol{r}) | \psi_i \rangle\right|^2 \delta\left(E_f - E_i - \hbar\omega\right),$$

where the perturbation potential $V = e\boldsymbol{A}(\boldsymbol{r}) \cdot \boldsymbol{p}/m_e$ with $\boldsymbol{A}(\boldsymbol{r}) = A_0 \boldsymbol{\epsilon} e^{i\boldsymbol{k}\cdot\boldsymbol{r}}$ and $\boldsymbol{p}$ being the momentum operator, $E_i$ and $E_f$ are the eigenenergies of the initial $|\psi_i\rangle$ and final $|\psi_f\rangle$ states, respectively. Given that the wavelength of the optical field is much larger than atomic dimensions and the typical lattice constant of LPCMO, the electric dipole approximation $e^{i\boldsymbol{k}\cdot\boldsymbol{r}} \approx 1 + i\boldsymbol{k} \cdot \boldsymbol{r}$ is applicable. Within this electric dipole approximation, the spectral weight is therefore governed by the dipole matrix element.

In LPCMO, the relevant low-energy optical absorption mainly involves $Mn^{3+}$ - $Mn^{4+}$ intersite charge-transfer excitations along the zig-zag chains. The transition matrix element is strongly controlled by the orientation of the occupied Mn $e_g$ orbital, the polarization of the optical field, and the oxygen-mediated Mn-O-Mn hopping pathway, as established by optical studies of spin/orbital-pattern-dependent absorption and optical anisotropy in manganites *(26,41,42)*. This highlights that optical spectral weight is highly sensitive to orbital orientation. In the CO/OO phase of LPCMO, the dominant in-plane orbital configuration favors $Mn^{3+}$-$Mn^{4+}$ charge-transfer excitations associated with the $d_{3x^2-r^2}/d_{3y^2-r^2}$ orbital texture, which reshapes the in-plane lattice potential and coherently drives the 2.45 THz mode. Such $Mn^{3+}$-$Mn^{4+}$ intersite charge-transfer/hopping excitations have been invoked in ultrafast studies of photoinduced dynamics in other manganites *(38-40)*. Upon entering the antiferromagnetic phase, the pronounced enhancement of the E // *c* conductivity indicates that the relevant low-energy charge-transfer transitions acquire a stronger out-of-plane character. This does not necessarily imply a substantial change of the static CO/OO ground-state orbital occupation. Instead, it suggests that the photoexcited $Mn^{3+}$-$Mn^{4+}$ charge-transfer final states contain an enhanced out-of-plane, $d_{3z^2-r^2}$-like component under the antiferromagnetic background, consistent with the established sensitivity of optical spectral weight to orbital-dependent transition matrix elements in manganites *(41,42)*. Such orbital reweighting strengthens the coupling to a *c*-axis-sensitive lattice coordinate and activates the spin-coupled 2.5 THz coherent phonon mode. Furthermore, because optical charge-transfer transitions in manganites are constrained by spin conservation and spin-dependent hopping, their spectral weight is also sensitive to the magnetic background. Therefore, upon further cooling, the development of magnetic correlations or ferromagnetic phase coexistence within the CO/OO regime can further modify the spin-dependent $Mn^{3+}$-$Mn^{4+}$ charge-transfer landscape. Through spin-phonon coupling and exchange-striction effects, this increases the effective force constant of the $c$-axis-sensitive lattice coordinate associated with the out-of-plane $d_{3z^2-r^2}$-like photoexcited states, leading to the hardening of the spin-coupled mode from 2.5 THz to ~2.6 THz *(43-45)*. In contrast, the 2.45 THz mode remains nearly unchanged because it is primarily governed by the more robust CO/OO-related in-plane lattice potential.

**Acknowledgments**

We gratefully acknowledge the support from the Bernina beamline of SwissFEL and Materials Science beamline (X04SA) of the Swiss Light Source in Paul Scherrer Institute. We acknowledge financial support by SNSF, Grant No. 200021_124496. The work at Rutgers University was supported by the DOE under Grant No. DOE: DE-FG02-07ER46382.

**Author contributions:**

Conceptualization: XL, RM, PB, US

Software: XL, RM, HTL

Methodology: XL, JC, RM, US

Investigation: XL, MS, SWH, SZ, AC, MR, JL, DB, SS, EP, PM, LW, SWC, CJ, PB, HTL, US, RM

Resources: SWC

Visualization: XL, RM

Project administration: XL, RM

Writing – original draft: XL, RM, US

Writing – review & editing: XL, MS, SWH, SZ, AC, MR, JL, DB, SS, EP, PM, LW, SWC, CJ, PB, HTL, US, RM

**Competing interests:** The authors declare that they have no competing interests.

**Additional information:**

Supplementary Information is available for this paper.

All data files will be uploaded to Public Data Repository and will be published and cited here using digital object identifiers before publication. Correspondence and requests for materials should be addressed to X. Liu.

# Supplementary Materials for

## Selective coupling of coherent phonons to intertwined charge-orbital and magnetic orders in doped manganites

X. Liu* *et al.*

*Corresponding author. Email: xin.liu@tuwien.ac.at; roman.mankowsky@psi.ch

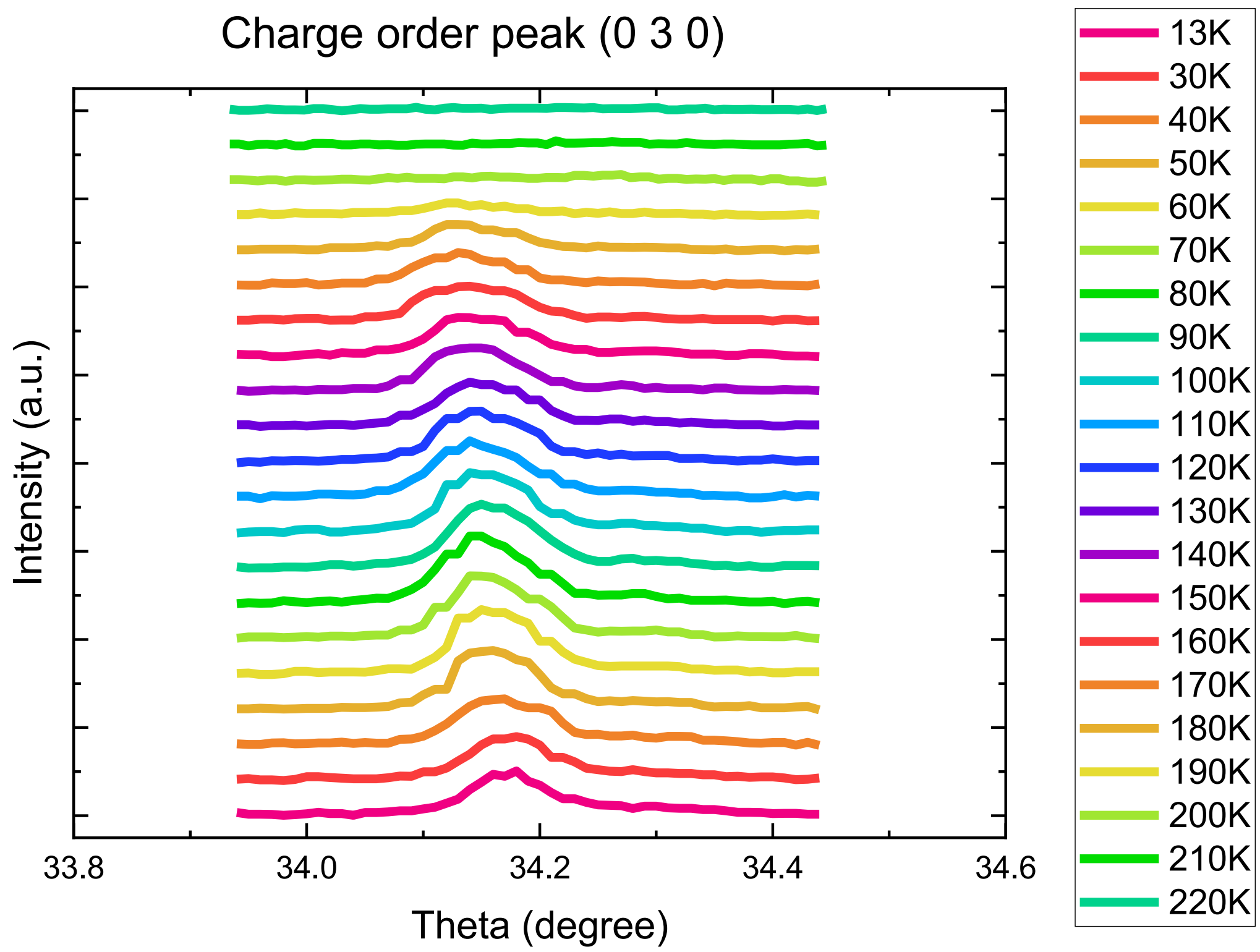

**Fig. S1.** Rocking curves of the (0 3 0) reflection taken at the Mn *K* edge at various temperatures.

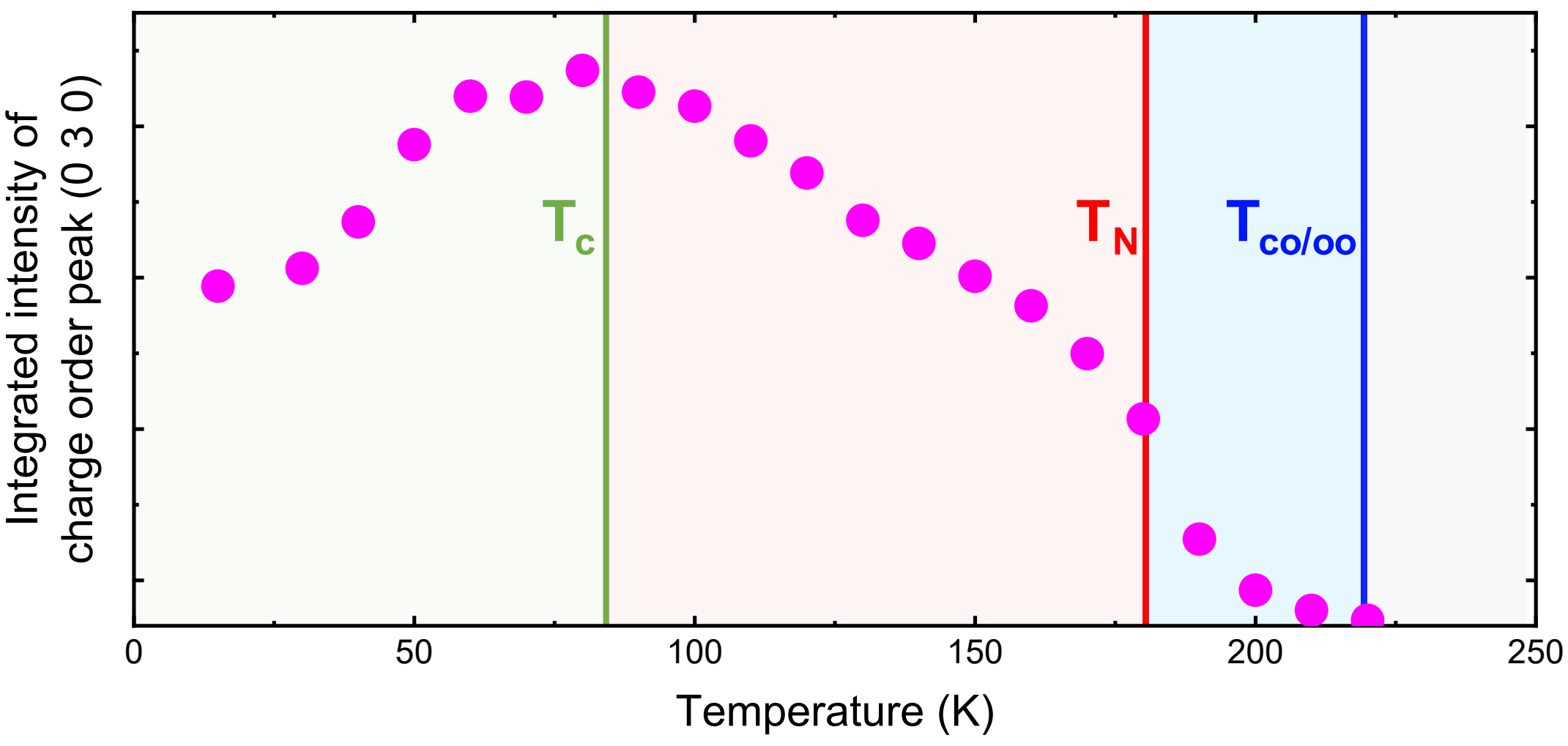

**Fig. S2.** Intensity of the charge order peak (0 3 0) as a function of temperature.

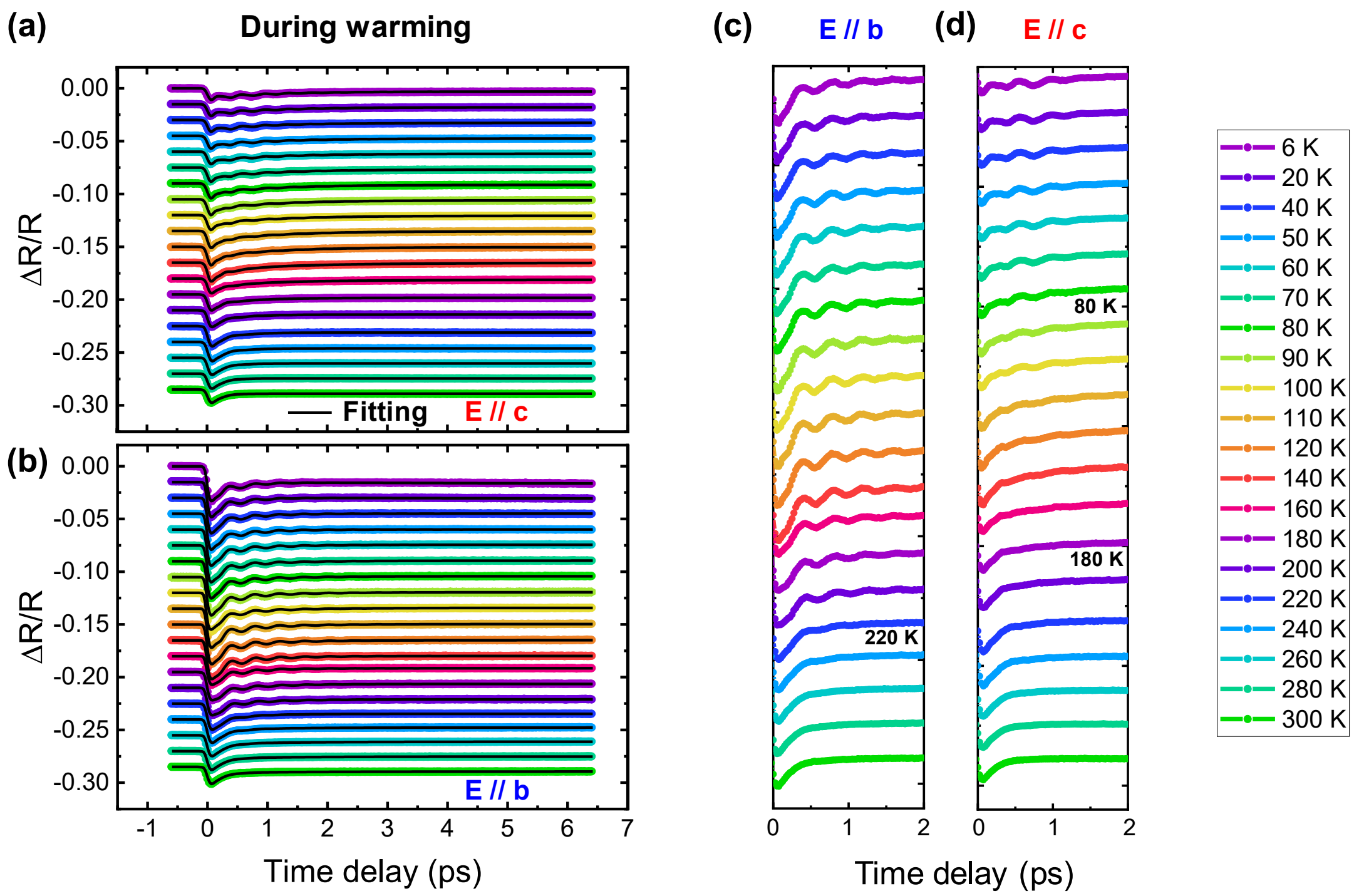

**Fig. S3.** Optical pump probe reflectivity as the function of temperature during warming. (a) Polarization of the probe beam along *c* axis (E // *c*). (b) Polarization of the probe beam along *b* axis (E // *b*). (c)(d) Zoom-in of figure (a) and (b).

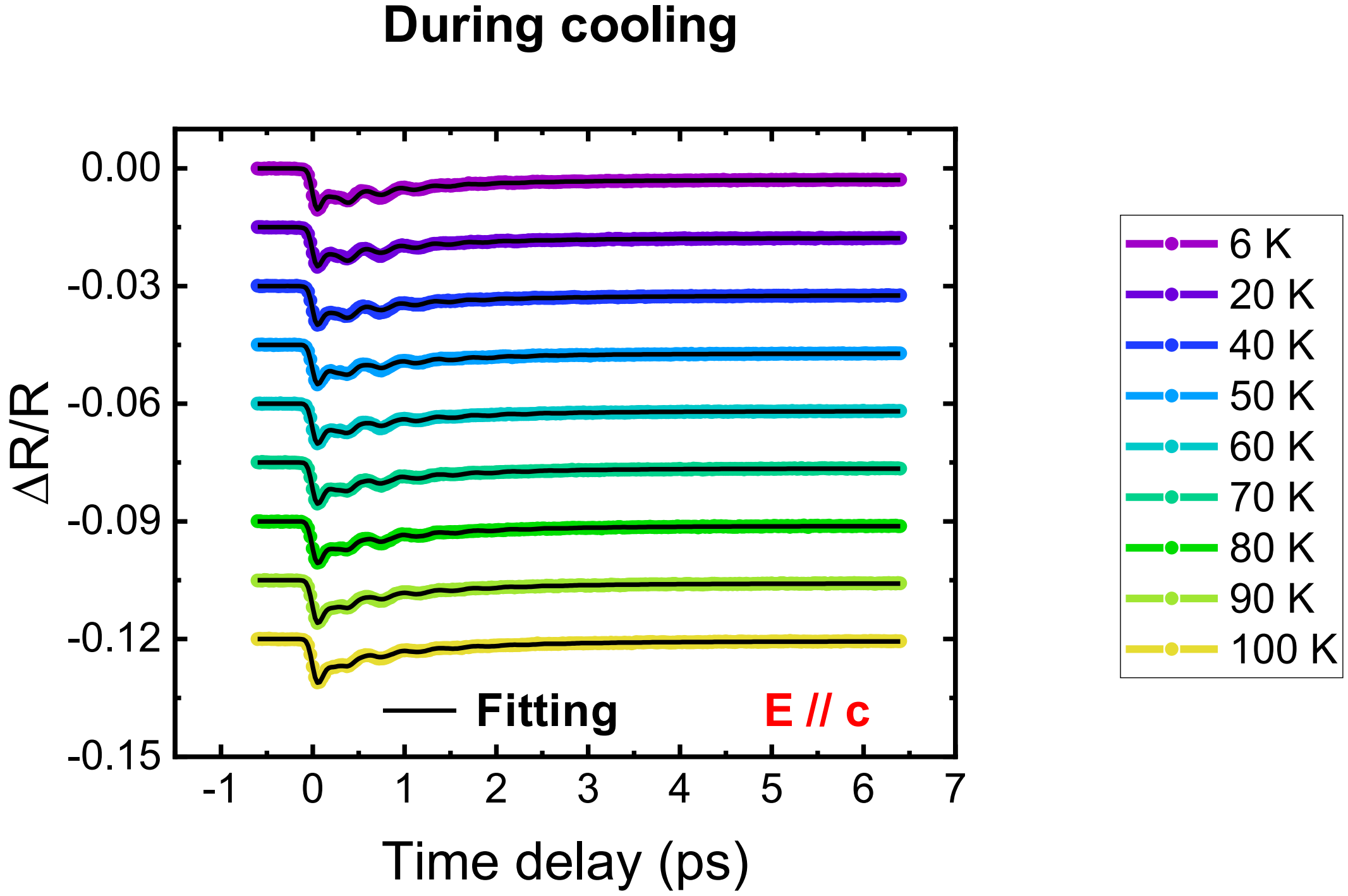

**Fig. S4.** Optical pump probe reflectivity as the function of temperature during cooling. The polarization of the probe beam is along *c* axis (E // *c*).

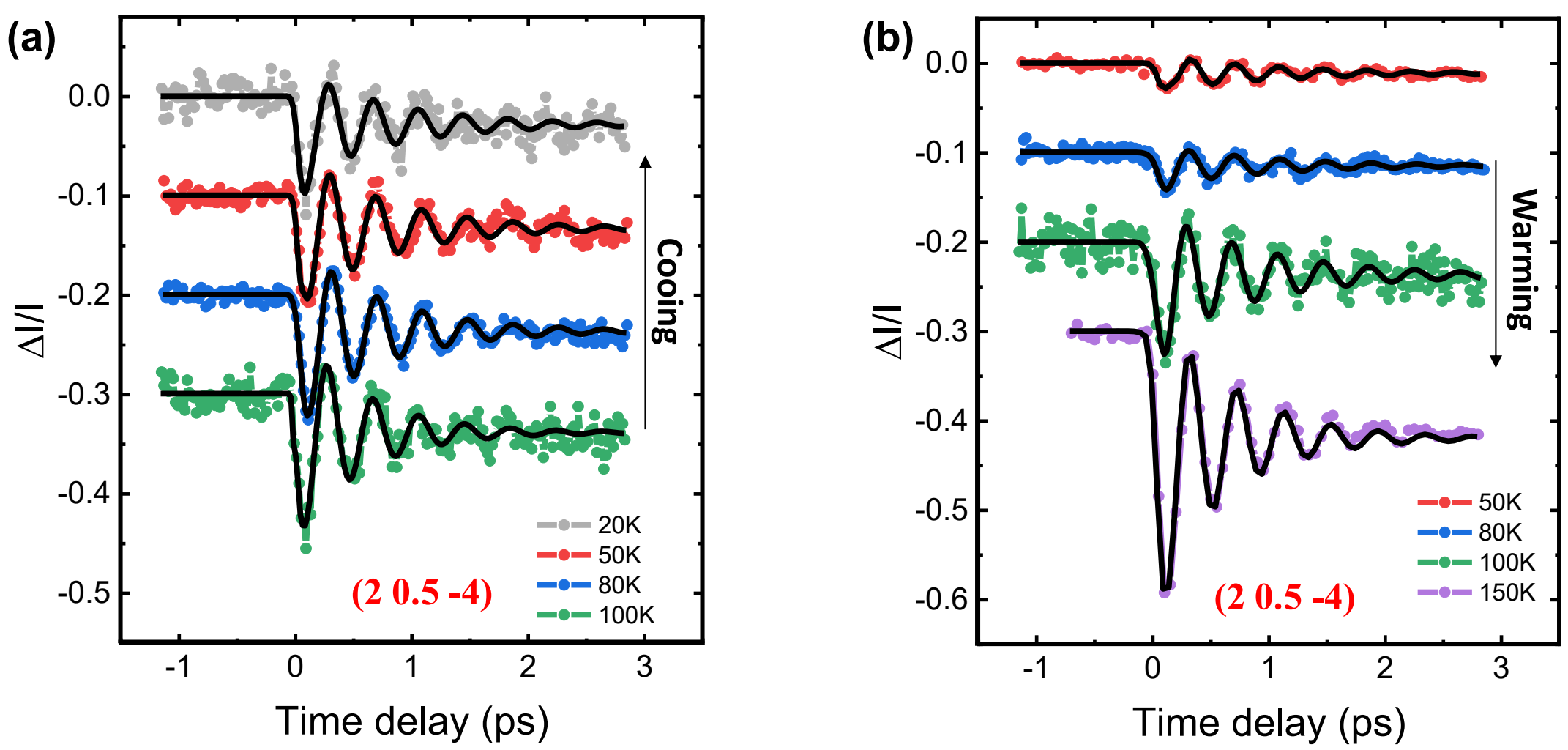

**Fig. S5.** Transient intensity changes of the (2 0.5 -4) reflection measured by time-resolved X-ray diffraction at various temperatures during cooling (a) and during warming (b).

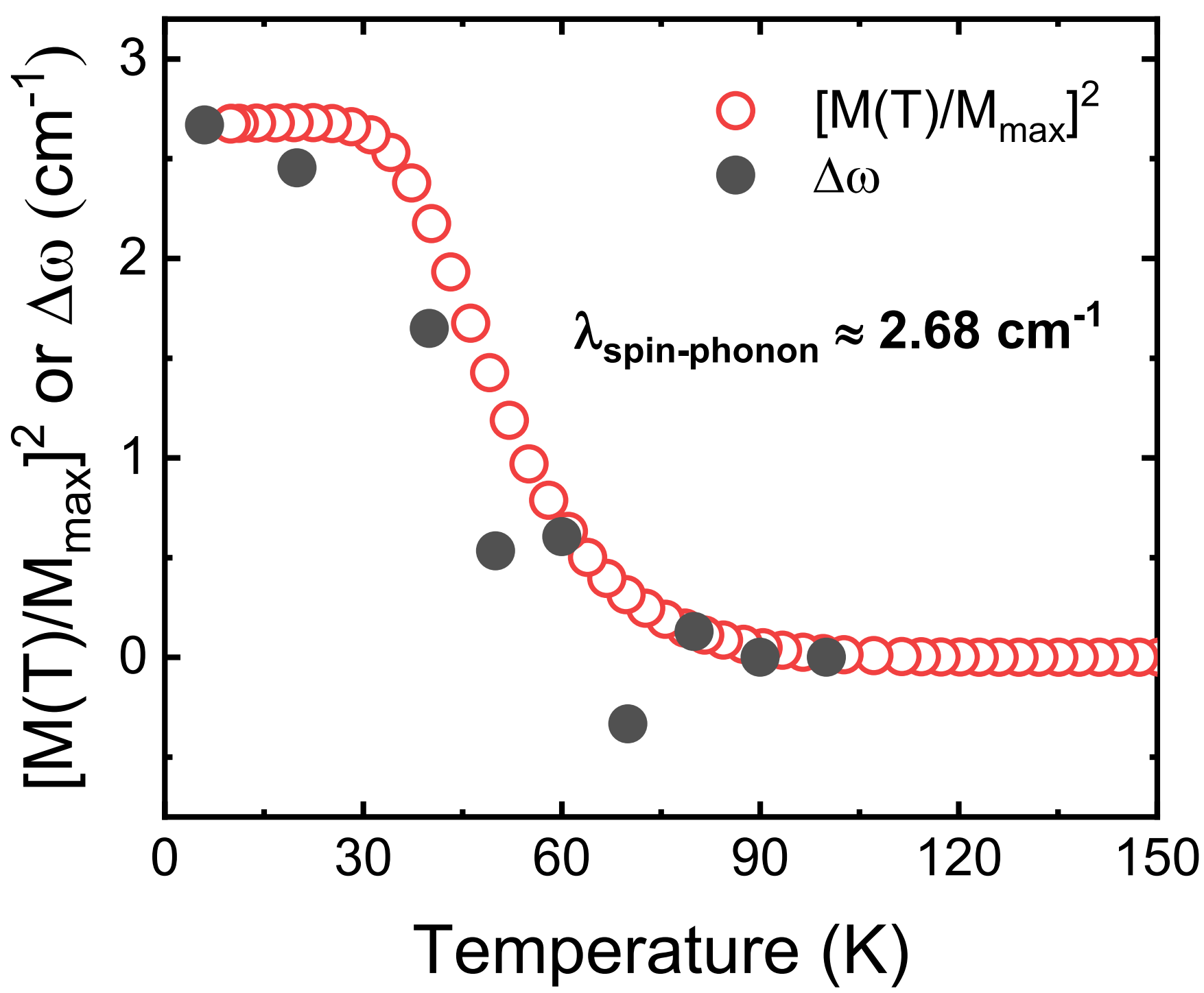

**Fig. S6.** Comparison of temperature dependent $\Delta\omega$ from optical pump probe reflectivity with $(M(T)/M_{max})^2$ from magnetization measurement. The spin-phonon coupling constant can be estimated using the following expression *(43,44)*: $\Delta\omega = \lambda_{spin-phonon} < S_i \cdot S_j >$, where $< S_i \cdot S_j > \sim (M(T)/M_{max})^2$. Therefore, $\lambda_{spin-phonon} = 2.68\ cm^{-1}$ can be obtained.

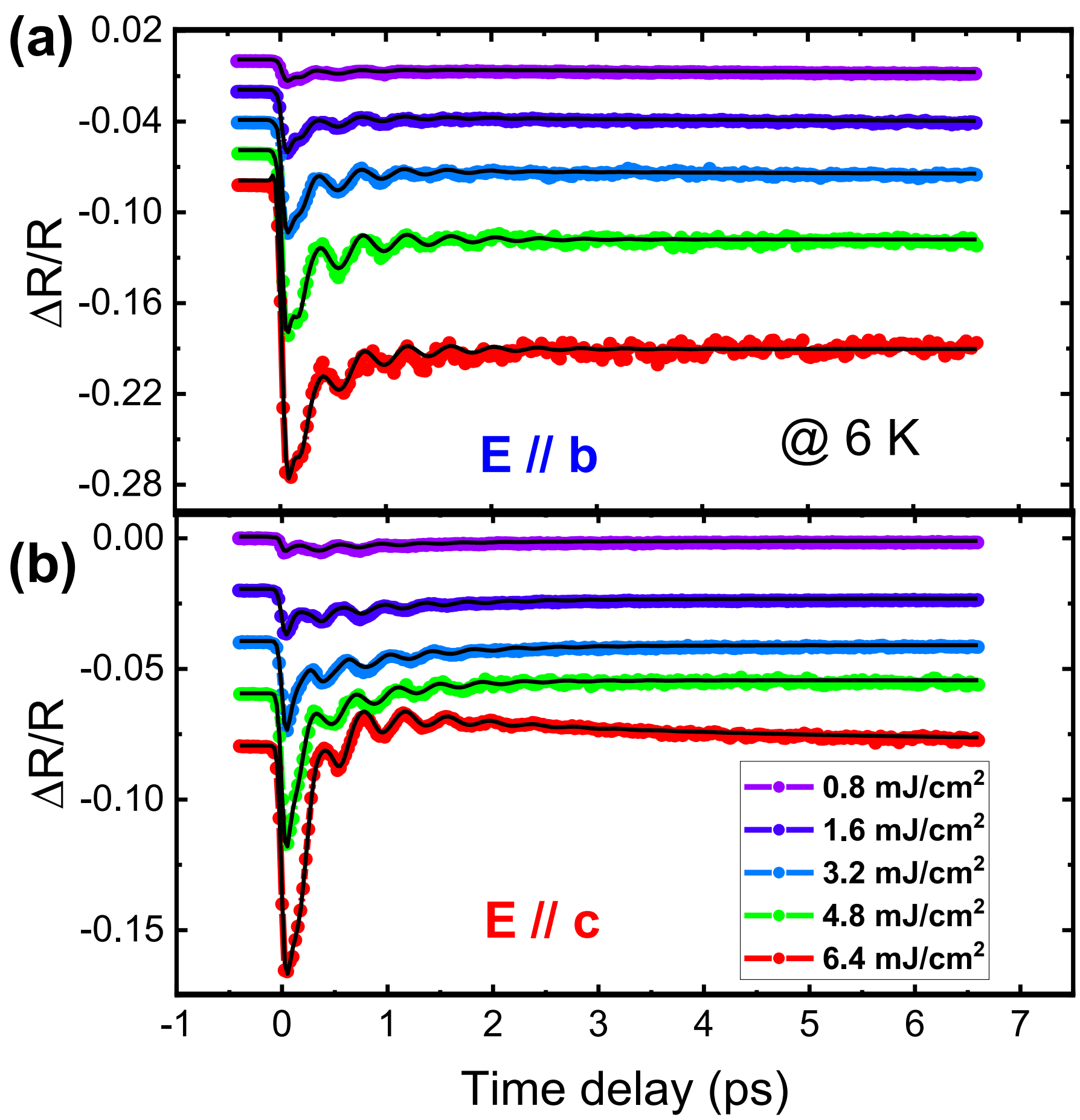

**Fig. S7.** Fluence dependent time traces of optical pump probe reflectivity for E // *b* (a) and E // *c* (b) at 6 K.

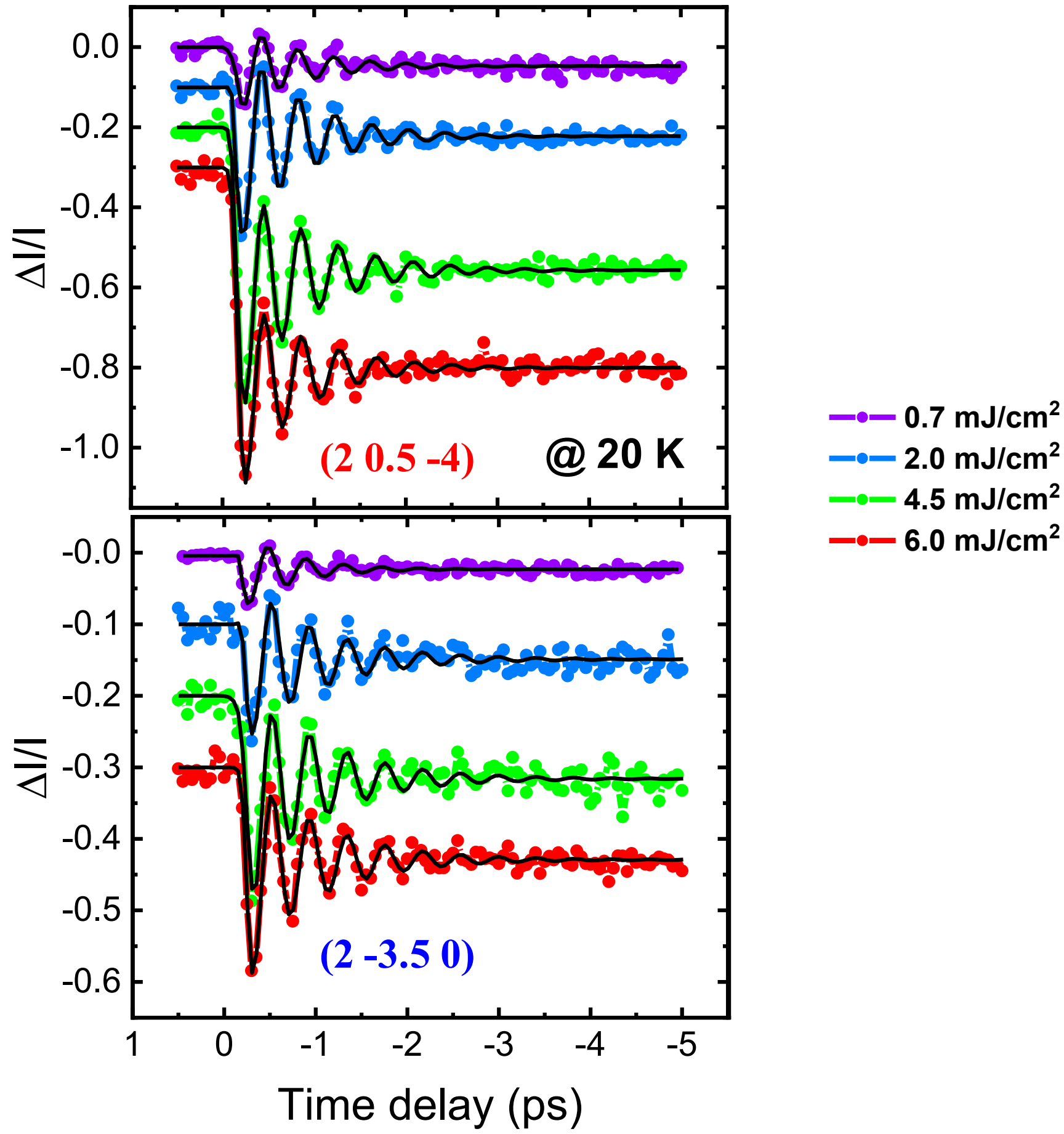

**Fig. S8.** Fluence dependent time traces of the transient intensity changes for the (2 0.5 -4) and (2 -3.5 0) peaks from X-ray diffraction taken at 20 K.

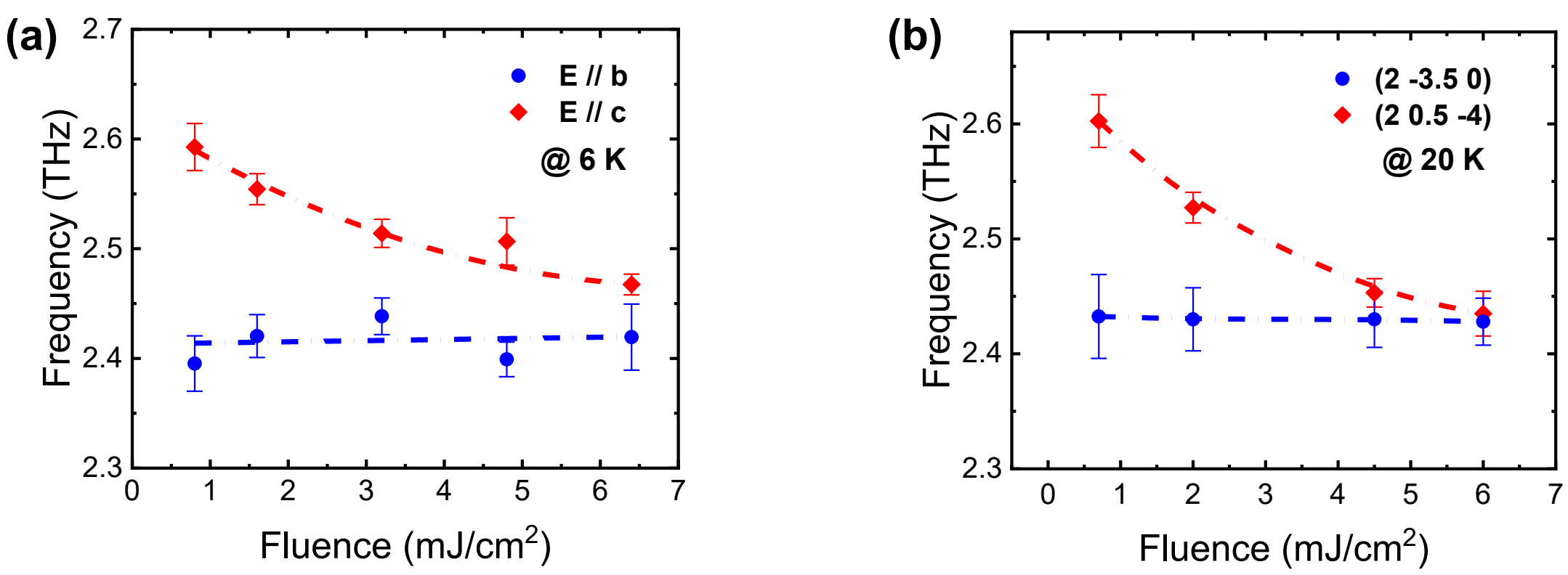

**Fig. S9.** (a) Fluence dependence of the frequency for E // *b* (blue circles) and E // *c* (red diamonds) at 6 K. (b) Fluence dependence of the frequency obtained from the (2 -3.5 0) (blue circles) and (2 0.5 -4) (red diamonds) superlattice peaks at 20 K.